%% file: main.tex
\documentclass[letterpaper]{article}
\usepackage{aaai24}
\usepackage{times}
\usepackage{helvet}
\usepackage{courier}
\usepackage[hyphens]{url}
\usepackage{graphicx}
\usepackage{natbib}
\usepackage{caption}
\usepackage{algorithm}
\usepackage{algorithmic}
\usepackage{newfloat}
\usepackage{listings}

\usepackage{tikz}
\usepackage{pgfplots}
\usepackage{array}
\usepackage{multirow}
\usepackage{fontawesome5}
\usetikzlibrary{backgrounds}

\usepackage[edges]{forest}
\usetikzlibrary{arrows.meta}

\usepackage{latexsym}
\usepackage[T1]{fontenc}
\usepackage[utf8]{inputenc}
\usepackage{microtype}
\usepackage{inconsolata}
\usepackage{csquotes}

\usepackage{url}
\usepackage{booktabs}
\usepackage{amsfonts}
\usepackage{nicefrac}
\usepackage{microtype}
\usepackage{xcolor}

\graphicspath{ {./images/} }
\usepackage{comment}

\usepackage{pgfplots}
\usepackage{pgfplotstable}
\usetikzlibrary{patterns}
\usepgfplotslibrary{colormaps}
\pgfplotsset{compat=1.18}
\usetikzlibrary{pgfplots.colormaps}
\usepackage{pgf-pie} 
\usepackage{xcolor,colortbl}
\usepackage{multirow}
\usepackage{xurl}
\usepackage{siunitx}

\usepackage{dcolumn}
\newcolumntype{d}[1]{D{.}{.}{#1}}

\usepackage{makecell}
\usetikzlibrary{matrix}
\usepackage{enumitem}

\newcolumntype{L}[1]{>{\raggedright\let\newline\\\arraybackslash\hspace{0pt}}m{#1}}
\newcolumntype{C}[1]{>{\centering\let\newline\\\arraybackslash\hspace{0pt}}m{#1}}
\newcolumntype{R}[1]{>{\raggedleft\let\newline\\\arraybackslash\hspace{0pt}}m{#1}}

\usepackage{booktabs}

\DeclareCaptionStyle{ruled}{labelfont=normalfont,labelsep=colon,strut=off} 
\floatstyle{ruled}
\newfloat{listing}{tb}{lst}{}
\floatname{listing}{Listing}

\title{The Implications of Open Generative Models in Human-Centered Data Science Work: A Case Study with Fact-Checking Organizations}

\author {
    Robert Wolfe, Tanushree Mitra
}
\affiliations {
    University of Washington\\
    rwolfe3@uw.edu, tmitra@uw.edu
}

\begin{document}

\maketitle

\input{sections/00_abstract}
\input{sections/01_introduction}
\input{sections/02_related_work}
\input{sections/03_approach}
\input{sections/04_data_pipelines}
\input{sections/05_motivations}
\input{sections/06_limitations}
\input{sections/07_discussion}
\input{sections/08_conclusion}

\bibliography{references}

\end{document}

%% file: sections/00_abstract.tex
\begin{abstract}

Calls to use open generative language models in academic research have highlighted the need for reproducibility and transparency in scientific research. However, the impact of generative AI extends well beyond academia, as corporations and public interest organizations have begun integrating these models into their data science pipelines. We expand this lens to include the impact of open models on \textit{organizations}, focusing specifically on fact-checking organizations, which use AI to observe and analyze large volumes of circulating misinformation, yet must also ensure the reproducibility and impartiality of their work. We wanted to understand where fact-checking organizations use open models in their data science pipelines; what motivates their use of open models or proprietary models; and how their use of open or proprietary models can inform research on the societal impact of generative AI. To answer these questions, we conducted an interview study with $N$=24 professionals at 20 fact-checking organizations on six continents. Based on these interviews, we offer a five-component conceptual model of where fact-checking organizations employ generative AI to support or automate parts of their data science pipeline, including Data Ingestion, Data Analysis, Data Retrieval, Data Delivery, and Data Sharing. We then provide taxonomies of fact-checking organizations' motivations for using open models and the limitations that prevent them for further adopting open models, finding that they prefer open models for Organizational Autonomy, Data Privacy and Ownership, Application Specificity, and Capability Transparency. However, they nonetheless use proprietary models due to perceived advantages in Performance, Usability, and Safety, as well as Opportunity Costs related to participation in emerging generative AI ecosystems. Finally, we propose a research agenda to address limitations of both open and proprietary models. Our research provides novel perspective on open models in data-driven organizations.

\end{abstract}

%% file: sections/01_introduction.tex
\section{Introduction}

Generative AI models have rapidly become a component of organizational infrastructure, with more than 90\% of Fortune 500 companies now using ChatGPT \cite{porter2023chatgpt}. Such models promise to transform information work by providing approachable conversational interfaces for performing complex tasks involving large quantities of text and data \cite{hassani2023role,eloundou2023gpts}. Recent research indicates that organizational integration of generative AI can complement the skills of educated professionals, especially early in their careers, increasing productivity and job satisfaction by automating repetitive tasks and making know-how of experienced workers more available to entry-level staff \cite{noy2023experimental,brynjolfsson2023generative}.

Despite this potential for positive impact, however, many scholars have voiced concerns over the growing reliance on closed, proprietary models \cite{birhane2022values}. Concerns have arisen from scholars in both Natural Language Processing (NLP) and the social sciences \cite{rogers-etal-2023-closed, ollion2023chatgpt}, responding to a growing body of research contending that ChatGPT and similar proprietary models can be used as a substitute for human subjects, both for labeling data in scientific studies \cite{Gilardi2023ChatGPTOC,amin2023will}, and simulating the behavior of human subjects \cite{park2023generative,shanahan2023role}, ignoring the paucity of technical information available about proprietary models and the uncertain reproducibility of results obtained. \citet{palmer2023using} contend that academic researchers should prioritize the use of \textit{open models} for which the weights are available for download, and training data is specified to the end user, unless they can provide an explicit, study-specific justification for choosing a proprietary model (\textit{e.g.}, studying the impact of OpenAI's DALL-E models on artists due to their widespread adoption \cite{jiang2023ai}).

\begin{figure*}
    \centering
    \includegraphics[width=.93\textwidth]{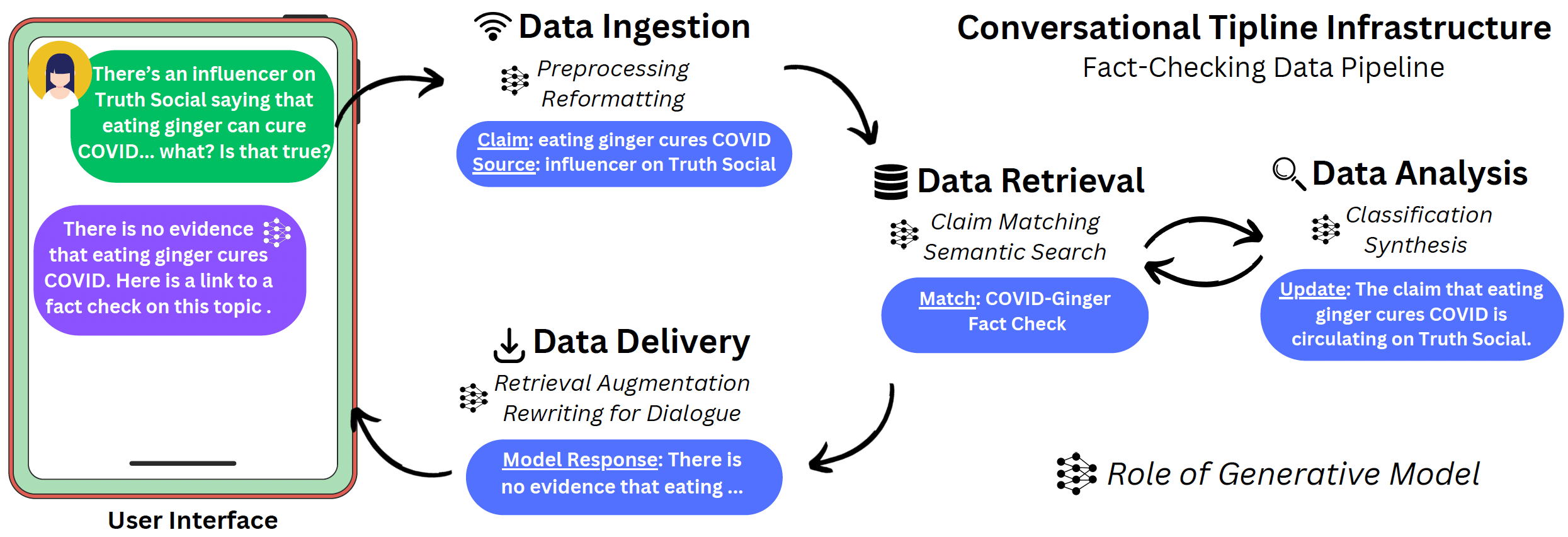}
    \caption{\small Conversational tiplines are novel data science pipelines for fact-checking accelerated by the advent of chat-based language models. Four components of conversational tiplines can leverage generative AI: Data Ingestion, Data Retrieval, Data Analysis, and Data Delivery.}
    \label{fig:tipline}
\end{figure*}

While these studies address the importance of open models for scientific integrity, they do not consider the impact that choosing open models can have on \textit{organizations} that adopt generative language models as technological infrastructure. In the present work, we seek to better understand the societal implications of open models by studying their use at fact-checking organizations, a group that shares several characteristics that render them worthy of consideration in this context. First, fact-checking organizations routinely employ state-of-the-art language models in their work, lest they find themselves overwhelmed by large volumes of misinformation \cite{das2023state,guo-etal-2022-survey}. Second, they must ensure the reproducibility, reliability, and impartiality of their work, or they will compromise both trust with their audiences and their membership in organizations such as the International Fact-Checking Network (IFCN) \cite{ifcnhome, walter2020fact}. And third, they play a vital role in maintaining the health of information ecosystems around the world \cite{li2023combating}. Understanding use of open models at fact-checking organizations can thus provide insight into the experiences of public interest organizations leveraging generative AI in an impactful sociotechnical context. In this work, we address three research questions:

\begin{itemize}
    \item \textbf{RQ1}: Where do fact-checking organizations employ generative AI models in their data science pipelines?
    \item \textbf{RQ2}: What motivates the adoption of open models by fact-checking organizations?
    \item \textbf{RQ3}: What prevents fact-checking organizations from further employing open models in their work?
\end{itemize}

\noindent To answer these questions, we conducted an interview study with $N$=24 professionals working at 20 fact-checking organizations across six continents. 
Adopting a human-centered approach to contextualize fact-checker perspectives on generative AI within the context of its use by practitioners, we found that fact-checking organizations reported employing generative models for Data Ingestion, Data Analysis, Data Retrieval, Data Delivery, and Data Sharing. Most participants preferred open models over proprietary models due to concerns related to organizational autonomy, data privacy and ownership, application and domain specificity, and model capability transparency. However, with a few exceptions, their use of open generative models was largely aspirational, as participants cited significant perceived shortcomings in the performance, usability, and safety of open models, as well as opportunity costs associated with not participating in emerging generative AI ecosystems offered by companies like OpenAI and Google. We make three contributions:

\begin{itemize}
    \item \textbf{We offer a five-component conceptual model to describe where fact-checker organizations employ generative models in sociotechnical fact-checking data science pipelines.} We offer two concrete examples of in-use pipelines that employ generative models: media monitoring pipelines, and retrieval-augmented conversational tiplines (illustrated in Figure \ref{fig:tipline}, the latter of which has seen significant improvements since the advent of general-purpose conversational models such as ChatGPT.
    \item \textbf{We offer taxonomies of 1) motivations of fact-checking organizations for preferring open models, and 2) limitations that prevent further adoption of open models.} We contextualize motivations and limitations by identifying the components of the data science pipeline wherein participants most located their impact, providing a grounded view of the relationship between the model itself and its organizational and societal impacts.
    \item \textbf{We propose a research agenda for addressing the concerns of fact-checking organizations with both open and proprietary generative models.} We offer concrete suggestions for research addressing the performance, usability, and safety of open models, which we suggest can help further their adoption. Given that general-purpose performance of proprietary models will mostly exceed open models, and the revenue of fact-checking organizations may be dependent on producing custom models that integrate with proprietary ecosystems, we also offer directions for research addressing transparency, agency, privacy, and specificity in proprietary models.
\end{itemize}

\noindent Rather than offering a prescriptive approach to open models or an empirical study of their effectiveness, we contribute an understanding of open models in a consequential setting, including how open and proprietary models are valued in practice. We believe these insights can inform the perspectives and research agenda of the AI ethics community.

%% file: sections/02_related_work.tex
\section{Related Work}

We review related work on open models, fact-checking organizations, and human-centered approaches to data science.

\subsection{Open and Proprietary Models}

While scholars have observed a spectrum of relative openness in AI releases \cite{solaiman2023gradient,langenkamp2022open}, we adopt the definition of open models proposed by \citet{palmer2023using} and \citet{rogers-etal-2023-closed}. Specifically, open models can be: 1) downloaded locally; 2) run without a call to an API; 3) shared with others; and 4) versioned. The contents of open models' training data must be disclosed, even if the data itself is not available \cite{palmer2023using}. This definition is satisfied by recent releases of generative language models such as Meta's LLaMA-2 \cite{touvron2023llama2} and its finetuned variants such as Stable Vicuna \cite{vicuna2023}. Some generative models, such as the popular Mistral-7B, are better characterized as ``open weight'' models \cite{jiang2023mistral}, as they make the model's weights available, but do not disclose training data to preserve competitive advantage. We note while many transformer-based models, such as Google's BERT \cite{devlin-etal-2019-bert}, would qualify as open models under these definitions, we study specifically \textit{generative} models.

The dominance of OpenAI's ChatGPT \cite{openai2022chatgpt} has engendered studies comparing its performance against that of finetuned open models. \citet{kocon2023chatgpt} find that ChatGPT's few-shot performance lags that of task-specific models such as fine-tuned transformers. Moreover, \citet{wolfe2024laboratory} find that small, open generative models can match or exceed ChatGPT after fine-tuning, while \citet{thalken2023modeling} show that fine-tuned Legal-BERT outperforms generalist models on classifying legal reasoning. While open generative chatbot models without fine-tuning have not matched the performance of GPT-4, which regularly tops evaluation leaderboards such as Stanford's Holistic Evaluation of Language Models (HELM) \cite{liang2023holistic}, recent open-weight models such as 01.ai's Yi \cite{ai2024yi} and Mistral AI's Mixtral 8x7B \cite{Jiang2024MixtralOE} have surpassed proprietary competitors including Anthropic's Claude 2 \cite{claude2} and OpenAI's GPT-3.5-Turbo \cite{openaimodels}, indicating that the best open chat models are comparable to proprietary models six months earlier. Our interviews show that perceived performance shortcomings of open models play a role in fact-checking organizations' use of proprietary models.

Most recently, proprietary model providers have allowed users to create customized versions of generative models, incorporating specialized prompts to govern behavior, and equipping models with external sources of data and programmatic actions. For example, OpenAI refers to customized versions of ChatGPT as ``GPTs'' \cite{openai2023gpts}, and allows ChatGPT Premium subscribers to build and share GPTs via OpenAI's ``GPT Store'' web platform, through which OpenAI has promised that builders of custom GPTs will be able to generate revenue \cite{openai2024gptstore}. The GPT Store is an example of an emerging AI \textit{ecosystem}, wherein users select from a range of generative models based on their needs. Our work offers insight into how fact-checkers foresee such AI ecosystems impacting their financial models. 

\subsection{Sociotechnical Infrastructure of Fact-Checking}

Fact-checking refers to assessing of the veracity of ostensibly factual information circulating in an information ecosystem that has the potential to harmfully mislead its audience \cite{graves2017anatomy}. While a version of fact-checking has long existed in the form of investigative journalism \cite{dickeyhistory}, modern fact-checking coincides with the rise of the internet and social media in particular \cite{graves2019fact}, which provided new conduits for the spread of misinforming content among vast networks of people. Fact-checking is typically described as a sociotechnical task \cite{graves2018factsheet}, meaning in this case that it is successfully accomplished by human investigators working with technological tools that support their expertise \cite{chopra2018sociotechnical, radiya2023sociotechnical}. Fact-checking organizations mostly embrace language models as essential infrastructure, expressing optimism about their potential to reduce manual workloads and exposure to harmful content, though they remain skeptical of AI that promises full automation of work that relies on human expertise \cite{juneja2022human}. Many modern fact-checkers adhere to the principles set forth by the International Fact Checking Network (IFCN) \cite{ifcnhome}, an organization created to establish common standards for fact-checking \cite{ifcnsignatories}. We interview professionals working primarily at signatory organizations of the IFCN to understand how they perceive the merits of open and proprietary models.

\subsection{Human-Centered Data Science}

As noted by \citet{berman2023}, interaction between data practitioners and the tools they use constitutes a social context that shapes the ethics of AI practices in organizations. Much work in social computing seeks to describe these tools and in interaction with human practitioners in data science \textit{pipelines} \cite{wang2019human}. For example, in a study of 183 data scientists, \citet{zhang2020} describe a common data science pipeline consisting of three high-level steps, including preparation, modeling, and deployment. However, as described by \citet{hopkins2021}, studies of data practitioners typically center on participants from big tech and academia, and may overlook the challenges faced by smaller organizations facing resource constraints, such as tensions between user privacy and organizational growth, a finding also echoed in \citet{bessen2022cost}, who note that AI startups may face tradeoffs between building more competitive and more ethical products. Human-centered data science centers the context in which data practitioners perform their work, acknowledging that data work work may be undertaken by domain experts or other workers not traditionally considered data scientists \cite{muller2019human}, a perspective that can yield domain-specific understandings of data pipelines. For example, \citet{Rothschild2022InterrogatingDW} note that civic workers at public and non-profit institutions exhibit skill with data contextualization that provides value far in excess of their sometimes less-developed computational abilities. 

Adopting a human-centered approach is well-motivated for fact checking, because despite the many benchmarks and techniques to support detection of misinformation \cite{russo-etal-2023-benchmarking,raj2023true,choi2023automated}, fact-checking organizations often view academic research as too detached from the real world \cite{juneja2022human,wolfe2024impact}, and recent work argues that even core NLP research on fact-checking should also study human factors \cite{das2023state}. We privilege the views of fact-checking professionals, surfacing where generative models fit into fact-checking data pipelines, and contextualizing the value fact-checking organizations see in open and proprietary models within those pipelines.

%% file: sections/03_approach.tex
\section{Approach}

We conducted an interview study with $N$=24 professionals at the 20 organizations shown in Table \ref{tab:orgs} to better understand the use of open models at fact-checking organizations. The study was approved by our university's IRB.

\subsection{Participants}

We reached out to 92 organizations via cold email, explaining our research and asking for an interview. We employed primarily purposive sampling \cite{etikan2016comparison} in emailing member organizations of the International Fact Checking Network (IFCN) \cite{ifcnsignatories} and their partner organizations, and snowball sampling \cite{naderifar2017snowball} when individuals at these organizations offered to connect us with another organization well-positioned for participation. Five participants at five organizations enrolled in our study as a result of snowball sampling; the rest enrolled as a result of purposive sampling. Individuals at ten additional fact-checking organizations responded to our emails but lacked technical knowledge needed to respond to our questions about open and proprietary models, as their roles were related to editing or upper management. We thus excluded them from this study. Most participants were engineers, research scientists, or department managers, with experience ranging from two years to 18 years in their current role. We refer to participants using a randomly assigned number between 1 and $N$ (\textit{e.g.,} P24 said ``\dots '').  \looseness=-1

\begin{table}[htbp]
    \small
    \begin{tabular}{|C{.95\linewidth}|}
    \toprule
    \multicolumn{1}{|c|}{20 Organizations in 15 Countries on 6 Continents} \\
    \midrule
   \multicolumn{1}{|c|}{\includegraphics[height=28mm]{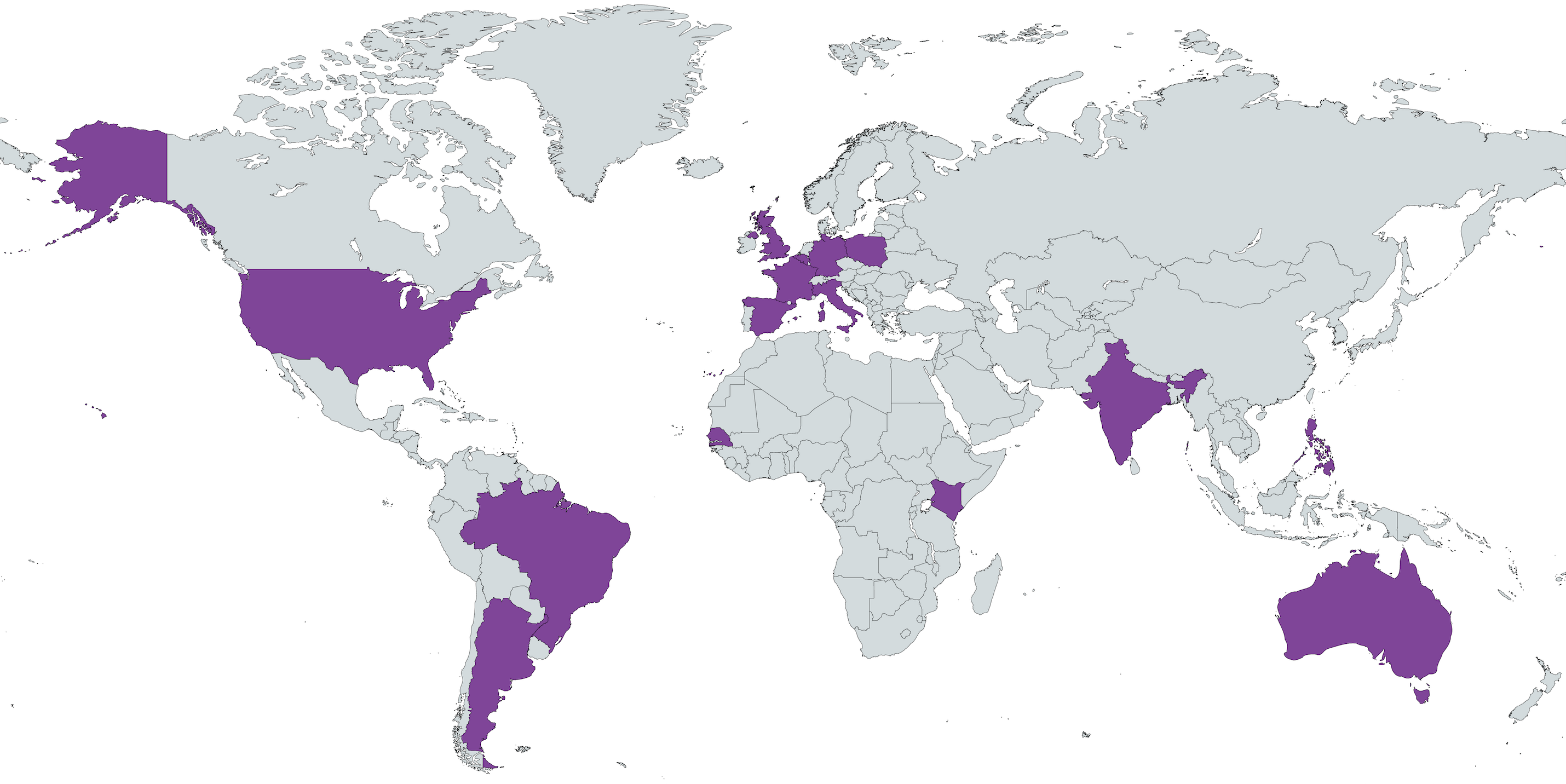}} \\
   \hline
    Australian Associated Press \cite{aap}, Africa Check \cite{africacheck}, Aos Fatos \cite{aosfatos}, Chequeado \cite{chequeado}, Code for Africa \cite{codeforafrica}, Der Spiegel \cite{derspiegel}, Factly \cite{factly}, India Today \cite{indiatoday}, Lead Stories \cite{leadstories}, logically.ai \cite{logically}, Maldita.es \cite{maldita}, Meedan \cite{meedan}, MindaNews \cite{mindanews}, Newtral \cite{newtral}, Pagella Politica \cite{pagella}, Pravda \cite{pravda}, Rappler \cite{rappler}, Science Feedback \cite{scifeedback}, Snopes \cite{snopes} \\
    \hline
    \end{tabular}
    \caption{\small Participants in 6 continents, 15 countries, 20 organizations.}
    \label{tab:orgs}
\end{table}

\subsection{Interview Process}

We developed a semi-structured interview protocol that asked participants about their organization's use of generative language models; the opportunities and challenges of generative AI in fact-checking; their organization's use of open models, and their motivations for adopting them; their reasons for using proprietary models; and what research could support the use of language models in fact-checking work. Where participants raised topics germane to the research but not covered in our interview protocol, we asked follow-up questions; for example, we asked P4 clarifying questions about their organization's beta release of a generative chatbot to collect misinformation circulating on platforms such as WhatsApp \cite{whatsapp}. Interviews lasted between twenty-five and ninety minutes and averaged approximately forty-five minutes. All interviews were conducted in English. 

\subsection{Data Analysis}

Interviews were recorded over Zoom \cite{zoom}, transcribed using Rev Max AI \cite{rev}, and manually corrected as necessary by the first author. To answer the study's research questions, we adopted a deductive-inductive approach to coding the interview transcripts. We employed the following deductive codes: Uses of Open Models, Motivations for Using Open Models, Limitations of Open Models, Motivations for Using Proprietary Models, and Implications for Research. The authors first coded four transcripts, and the first author created an initial codebook that included inductively generated themes. The last author reviewed the codebook and the authors jointly revised the codes. The authors then coded four additional transcripts at a time until all transcripts had been coded, reviewing and revising the codebook after each round of coding. Finally, the authors followed a thematic analysis process \cite{Braun2022EverythingCW} to generate themes that answered the study's research questions, using shared memos to precisely define the themes. 

%% file: sections/04_data_pipelines.tex
\begin{figure*}
    \centering
    \includegraphics[width=\textwidth]{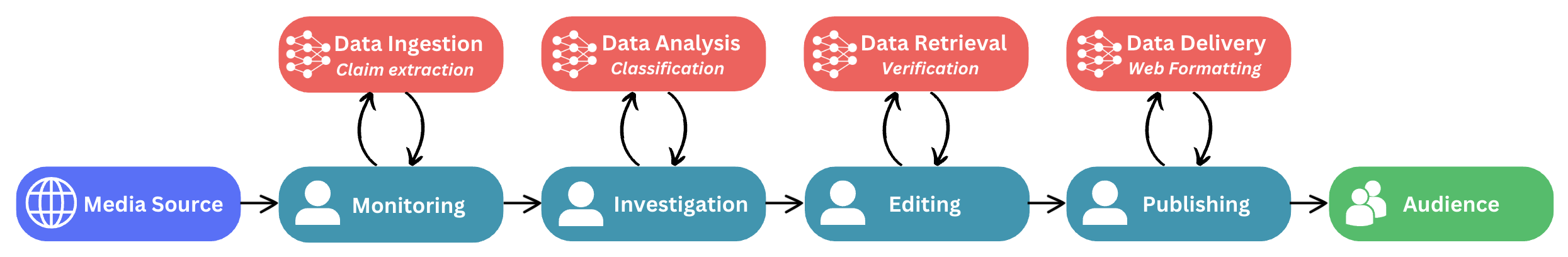}
    \caption{\small A sociotechnical media monitoring fact-checking pipeline, with generative AI in red, and human fact-checking processes in teal.}
    \label{fig:media}
\end{figure*}

\section{Data Pipelines in Fact-Checking}\label{sec:findings}

Consistent with prior work in human-centered data science, we found during thematic analysis that understanding the motivations of fact-checking organizations for using open or proprietary models requires understanding the ways in which they collect, analyze, and exchange fact-checking data - and specifically where they use generative models in these processes. To that end, we begin by providing a conceptual model of five components of fact-checking data science pipelines in which participants described using generative models. We assign each component an icon subsequently used in the Motivations and Limitations sections to associate participant perspectives with components of the pipeline.

\subsection{\faRss \ Data Ingestion}

Participants reported using generative AI to collect and preprocess data, whether via \textit{media monitoring} efforts employing AI-driven tools designed by social media companies or by the organizations themselves, or via \textit{tipline interfaces} wherein a user can submit misinformation for fact-checking. We refer to this component of the data science pipeline as \faRss \ \textbf{Data Ingestion}, and denote it using an RSS icon \faRss \ to suggest the role of monitoring novel information. \looseness=-1

Most participants described media monitoring pipelines like that illustrated in Figure \ref{fig:media} as an essential means of observing circulating misinformation. P3 noted that ``social media listening is the main point of entry'' to their data pipeline,  while P11 said that they focus on monitoring WhatsApp because ``the coverage is so massive'' in their country. P17 said that automated approaches including generative AI were necessary for media monitoring ``given the volume of production and how much content we can reasonably digest.'' P2 noted that one of their primary uses of generative AI was automating the data ingestion stages of their data pipeline: ``we're totally focused on this pipeline, and we're capable of automating the monitoring and the detection phases.'' Participants also used generative AI to preprocess data for other pipeline stages. P16 said that they use generative AI to ``clean up data'' and that ''those things [language models] are a time saver'' when ingesting data. P19 described using generative AI such that ``content can be synthesized and reformatted'' for analysis. \looseness=-1

Participants also described gathering information using conversational tiplines (\textit{i.e.,} chatbot interfaces) via which audiences can submit potential misinformation. Such interfaces are novel in comparison to media monitoring: P24 described an internal conversational interface that existed as early as 2016, but this interface was never used for data collection. P11 said they first developed a user-facing tipline in 2019, and that they had to significantly scale the tipline during the COVID-19 pandemic, as user interactions increased more than tenfold. Conversational models utilizing modern generative AI first saw release in 2023, as P4 released a beta for a tipline utilizing ChatGPT for its back end, and P11 improved their existing framework with generative models. Generative tiplines serve several purposes, including engaging audiences and bringing in data from sources not easily observed through media monitoring. P24 noted conversational tiplines help to observe ``especially provincial media that we are not that aware of or that we're not looking at regularly,'' while P14 noted they use a tipline to bring in new claims for investigation. P20 expressed interest in adapting their existing tipline to utilize generative AI for ``constraining the conversation with [the user] to elicit more data that's actionable.''

\subsection{\faSearchPlus \ Data Analysis}

Participants reported using generative AI to analyze large volumes of potentially misinforming content, leveraging its capabilities to parse highly contextual information, and utilizing few-shot approaches to avoid fine-tuning additional models. We denote \faSearchPlus \ \textbf{Data Analysis} with a magnifying glass \faSearchPlus \ to suggest the closer study of information. \looseness=-1

Participants reported using generative AI to support \textit{classification} and \textit{synthesis} of text and multimodal content. P4 used generative AI to classify text based on constructs like urgency that they previously captured using proxies like message formatting: ``instead of counting how many exclamation points a post has or how many caps it uses\dots Gen AI\dots give[s] us a score from zero to a hundred of the sense of urgency.'' P15 noted using generative models as part of an ensemble of deep learning and rule-based NLP classifiers. P2 used generative AI to extract and classify ``patterns of manipulative messages\dots like the government doesn't want you to know this or share it,'' noting that generative AI can be ``something like an anti-spam filter'' for misinformation. P3 said they use generative AI for classifying multimodal misinformation, including memes: ``the text itself isn't disinformation. The image without the text isn't disinformation. The image plus the text can feed very clearly into a disinformation narrative.'' P2 noted that user-friendly generative AI enables less technical fact checkers to create classifiers: ``Generative AI\dots democratizes who can work with AI\dots with an API and a little magic with a prompt, you can have something really powerful.'' \looseness=-1

Participants also embraced uses of generative AI for synthesizing data. P22 said that ``one of the values of generative AI is really synthesis, and looking and combing through tons of material, which\dots [we] will not have time for.'' P1 noted using ChatGPT to synthesize hundreds of documents collected every day via media monitoring, and to structure the data ``in a tabular format\dots in 90\% of the cases, it gives me a nice table.'' P3 used generative AI to synthesize component claims into narratives, improving the scalability of fact-checking work: ``large language models are super good at basically clustering claims into a narrative. Fact checking just individual claims is whack-a-mole, a losing proposition\dots you'll never be able to scale it.'' P11 demoed a GPT-4 driven narrative system for us, explaining that it synthesizes new claims into overarching narratives already associated with fact checks, pending human review: ``this summary is linked to these four different contexts that we have already received\dots generative AI here has proposed to us something that is a bit overarching\dots [it] is seeing what we produce, and the evidence that accompanies [our] debunks, and is proposing to us [a fact check] that has already been verified by a human.''

\subsection{\faDatabase \ Data Retrieval}

Participants reported using generative AI to facilitate \faDatabase \ \textbf{Data Retrieval} from catalogues of past fact checks or other verified sources of information maintained by the organization. We denote Data Retrieval with a database \faDatabase \ to suggest the retrieval of stored data. \looseness=-1

Many participants described using \textit{Retrieval-Augmented Generation} (RAG) \cite{lewis2020retrieval} to allow GPT models to incorporate factual data. P4 described ``a RAG pipeline that connects OpenAI's GPT-4 with our database of fact checks\dots we have all the articles and fact checks that we ever published stored as embeddings. And then\dots we perform semantic search using cosine similarity, and we take the most relevant results.'' P19 described retrieving data from a catalogue of past fact checks to support fact-checker investigations. P13 noted using third-party reliability metrics to determine what external content can be accessed using RAG. P11 noted that RAG allowed them ``to go beyond very basic keyword searches\dots so that the search in the database was actually fruitful and accurate.'' Some participants adopted more complex methods. P22 described creating an internal knowledge graph from which generative models could retrieve content: ``We started building our own knowledge graph, our own ontology, and using that, the structured data from that, to generate content.'' P23 created custom GPTs to retrieve data: ``we built it on the claim review database\dots only our factcheck articles\dots so interacting with that search persona would give results only from the database along with a source link.'' P23 also produced custom models to retrieve external data: ``Parliament data is completely public, so we had the tech team\dots scrape the entire database \dots and the persona only picks up responses from this dataset.''

\subsection{\faDownload \ Data Delivery}

Participants reported using generative AI to support \faDownload \ \textbf{Data Delivery} to end users on websites or social media, as well as by providing automatic responses to users via conversational tiplines. We denote Data Delivery with a download icon \faDownload \ to suggest the transfer of data to the end user.

Among the most common uses of generative AI reported was to format content or generate metadata prior to sharing it with audiences. P1 described using GPT models to ``generate hashtags for\dots mini FactCheck videos that we publish on TikTok.'' P8 said that ``in most daily use cases, in terms of generative AI use, I would say it's help with promotion. So all of the social media content, coming up with summaries for SEO purposes for article publishing, title generation.'' P5 noted that they use a generative AI-backed tool that ``suggests times the best times for us to post our content on social media based on the type of demographic our subscribers are, or our audiences are, and when they're using social media.'' \looseness=-1

Participants who used conversational tiplines reported using generative AI to \textit{deliver} information to end users, in addition to \textit{ingesting} misinforming content. P4 said that incorporating ChatGPT into their tipline was ``an obvious use case to improve a product that was already relevant for our readers,'' noting their conversational tipline had ``over 70,000 users, which for an organization our size is quite a lot.'' Several participants described an evolving information ecosystem wherein users sought information from specialized conversational agents, rather than traditional search applications. P21 said that they see generative AI as ``a preferred medium for somebody to get at the work that we have done\dots it is summarizing or reporting on work that was done by the trusted fact checkers.'' P24 said that they use a WhatsApp chatbot that ``allows us to answer to a high volume of messages,'' noting that ``if you could actually ask the [chatbot], can you please tell me what the inflation was in the last five years? And it could actually answer you with information that comes from a reliable source, which we know is one of the big problems of generative AI, we think that's an enormous leap forward in the way that we can actually reach people with verified information.'' P19 contended that conversational agents can assist users ``even if we don't have a fact check\dots explain this persuasion technique that's being used or the trope that is being repeated.'' 
P9 envisioned reaching younger audiences with an in-progress tool allowing a user to ``chat with our archive\dots this is a hurdle for young people to get into the discourse, that some background knowledge is missing. Maybe the AI could help.'' \looseness=-1

\subsection{\faUsers \ Data Sharing}

Participants noted that generative AI assists in \faUsers \ \textbf{Data Sharing} between organizations, in that they structure data for sharing, and serve as shared computational infrastructure. We denote Data Sharing with a Users icon \faUsers \ suggesting transfer of data between organizations. \looseness=-1

P18 described generative AI as a tool for exchanging fact check data and collectively scaling audiences to address problematic information: ``especially during elections, we try at (anonymized) to bring together other fact checkers so that people are not working in their own little silos\dots that's one area in which generative AI can really help. If fact-checkers are working together, whatever data they have, they can help scale the impact of their fact check to different segmented audiences that they serve.'' P3 described a prototype generative AI system operating on shared data, noting that especially in the case of elections, ``fact checkers are banding together to offer a united response\dots European checkers will put their claims in a common database, and we will build systems in which, when we detect a new narrative\dots [if] it's present more than one or two countries, there will be a special task force that will be tasked with producing a debunk\dots we could not do it with a BERT or a Sentence BERT\dots early tests are promising with a [generative] large language model. This is also where the multilingual aspect comes in very handy.'' P11 described a similar multi-organization ''project where we have installed [generative] technology\dots at fact-checkers in all Russia bordering countries. And we are also looking at\dots Spanish speaking Latin America. And we can see how common threats appear\dots where there are common narratives that point to particular actors.''

%% file: sections/05_motivations.tex
\begin{figure*}
\small
\centering
\begin{forest}
  for tree={
    align=left,
    font=\sffamily,
    l sep'=8pt,
    s sep = 4pt,
    fork sep'=4pt,
    draw,
  },
  forked edges,
  baseline,
  where level=0{
    tikz={\draw [thick] (.children first) -- (.children last);},
  }{},
  [\textbf{Motivations for Open Models in Fact-Checking}, fill=gray!5
    [\textbf{Organization Autonomy}, fill=gray!5, align=center
        [\faSearchPlus \ Reliable Availability \\
        \faRss \ \ Intentional Design \\
        \faRss \ \ Avoiding Gatekeeping \\
        \faUsers \ Shaping the Future, fill=green!10
    ]    
    ]
    [\textbf{Data Privacy \& Ownership}, fill=gray!5, align=center
        [\faDatabase \ \ Protecting Internal Data \\
        \faUsers \ Publisher Solidarity \\
        \faDownload \ \ Legal Safety \\
        \faUsers \ Mitigating Shared Risk, fill=blue!10
        ]
    ]
    [\textbf{Application Specificity}, fill=gray!5, align=center
        [\faSearchPlus \ Task Performance \\
        \faSearchPlus \ Cost Efficiency \\
        \faDownload \ Domain-Specific Tone \\
        \faSearchPlus \ Toxic Content Control, fill=purple!10
    ]
    ]
    [\textbf{Capability Transparency}, fill=gray!5, align=center
        [\faSearchPlus \ \ Performance Transparency \\ \faDownload \ \ Avoiding Deceptive Marketing \\
        \faUsers \ Tools for Replication, fill=violet!10
    ]
  ]
  ]
\end{forest}

\hfill \break

    \caption{\small A taxonomy of the motivations of participants for preferring open models over proprietary models in fact-checking organizations.}
    \label{fig:motivations_tree}

\end{figure*}
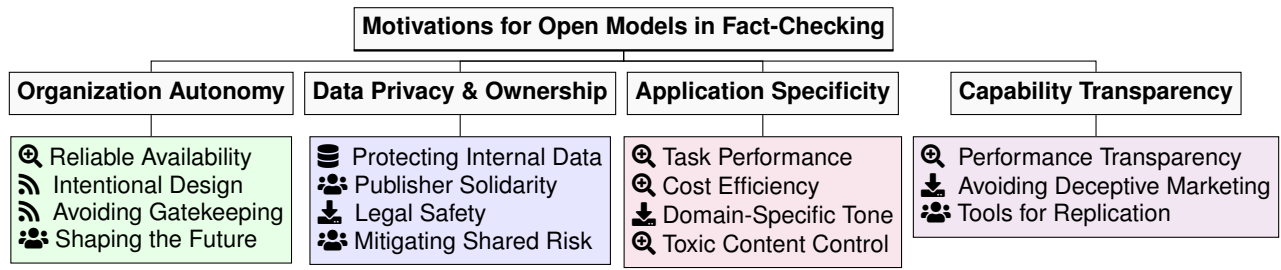

\section{Motivations for Open Models}\label{sec:motivations_open_models}

We found that four primary concerns motivate the use of open models: Organizational Autonomy, Data Privacy and Ownership, Task Specificity, and Capability Transparency. We describe these concerns in turn, making reference to the components of the data pipeline with which they intersect.

\subsection{Organizational Autonomy}\label{sec:autonomy}

Participants expressed concern that dependence on proprietary generative models could compromise the autonomy of their organizations. They noted that open models offer more \faSearchPlus \ \textbf{Reliable Availability} than proprietary models, which could be affected by unexpected deprecation or provider instability. Several participants described uncertainty about using OpenAI's models in the wake of its CEO's firing and reinstatement. P4 said ``the whole OpenAI drama was an eye opener. If OpenAI goes bankrupt tomorrow, then it's really bad to build products that depend on their software and on their API. We'd rather just host all the models that we use.'' P2 said that their organization is building next-generation content moderation and misinformation detection tools to circumvent this dependence, while P7 noted that their company builds on open models, mitigating issues of deprecation: ``We primarily rely on open source technologies. When Facebook releases their models and they open source it, that's what we use. We don't primarily rely on any [closed] corporate models.'' Participants also preferred open models because they facilitated \faRss \ \textbf{Intentional Design} of tools for fact-checking use cases. P8 said that proprietary tools offered by social media companies were often inadequate because they ``are targeting brands. And brands like McDonald's will have certain sets of keywords that will not change over time\dots we want to have this additional aspect, which is discovering new keywords because we want to stay on top of narratives.'' P24 said ``for example, social media monitoring or social listening tools\dots we end up developing a lot of things out of need\dots because we don't have the same needs as a marketing agency.'' Similarly, participants noted that \faRss \ \textbf{Avoiding Gatekeeping} by corporations motivates open models. P10 said that ``we have to prove ourselves to the social media companies'' to gain access to the only tools to monitor and address misinformation on their platforms, noting this process is onerous for resource-challenged local organizations. Finally, participants said open models could help fact-checking organizations in \faUsers \ \textbf{Shaping the Future} of fact-checking. P9 argued for accelerating adoption of open generative models: ``This all brings chances and holds a lot of potential for us, and we should be the one who shape this development, and not let others be the one who dictate how we have to deal with this at some point because we waited too long.'' Similarly, P22 remarked that ``fact-checkers need to be part of the creation of these tools\dots a lot of the tools that we are seeing don't really quite fit our use cases\dots it's about the process by which things are created.''  \looseness=-1

\subsection{Data Privacy and Ownership}\label{sec:data}

Participants described concerns surrounding data privacy and ownership as a central motivation for using open models. P9 preferred using open models for \faDatabase \ \textbf{Protecting of Internal Data}, saying that ``we have very sensitive material that we're working with here, investigative reporting and investigative stories, and we don't want this to be used in [corporate] models and as training material,'' and further noting that they use cloud instances hosted only in Europe for ``a sort of legal safety'' due to stricter European data protection laws. P4 said that, due to data privacy concerns, ``we have as a policy to always prioritize using open source software where we can.'' P21 noted ``I blocked [OpenAI's] crawlers from being able to train on our content until some type of commercial compensation becomes available. I think most of us are kind of waiting, holding our breaths for the New York Times case lawsuit to play out because otherwise individual orgs the size of the fact checkers, we don't really have the leverage to accomplish what that lawsuit stands to do in setting a precedent.'' Participants also said that \faUsers \ \textbf{Publisher Solidarity} motivated use of open models over proprietary models that profit from journalistic organizations' data without consent. P4 noted their discomfort with ``the notion that those companies are profiting using other companies' and other people's work.'' Participants also preferred open models that disclosed their training datasets, providing a sense of \faDownload \ \textbf{Legal Safety}, especially for user-facing applications. P4 said that ``the issue of copyright is a big one, especially for image generation\dots we could never use anything, any tool that generate images in our workflow, because we don't know how most models were trained.'' P8 noted that they delayed using generative models due to fears of copyright infringement: ``With generative AI, we were\dots scared to use it, because of the fact that we don't want to feel like we are plagiarizing\dots because of the possibility of\dots [copying] other articles.''  Participants noted that open models at least disclose their training datasets, offering some clarity concerning the risk of infringement. Finally, participants preferred open models for \faUsers \ \textbf{Mitigating Shared Risk} when building technologies for the fact-checking community. P4 created a transcription tool for the community using OpenAI's models, but noted is not in production because ``a lot of people have concerns about sending their data - interviews, important interviews - to OpenAI.'' \looseness=-1

\subsection{Application Specificity}\label{sec:specificity}

Participants preferred open models for specific applications that demanded high performance, domain-specific tone, and control over toxic content. P14 said that their organization preferred small, fine-tuned open models for internal data analysis, noting that they exhibit stronger \faSearchPlus \ \textbf{Task Performance} and \faSearchPlus \ \textbf{Cost Efficiency} than proprietary generative models. P6 echoed this, noting that their organization still prefers thoroughly vetted, task-specific models for many tasks, despite hype about replacing these methods with proprietary models. P10 considered GPT models one of many tools, not the sole solution to any problem involving language, despite the marketing of proprietary models. Participants also noted that proprietary models intended for general-purpose use couldn't achieve \faDownload \ \textbf{Domain-Specific Tone} for end user applications. P4 said ``out of the box, GPT-4, for instance, you get very decent results, but it's also very generic.'' P8 noted that, as a result of RLHF, GPT-4 ``sounds completely unnatural,'' rendering it difficult to incorporate in their user-facing applications. Participants also reported using open models to gain more granular \faSearchPlus \textbf{Toxic Content Control}. Participants including P17 took issue with corporate models' one-size-fits-all approach to alignment with human values, which they noted could hamper the model's ability to respond to toxic and hateful content that fact-checking organizations handle during their work. P5 noted that the OpenAI's RLHF process makes it difficult for models ``to unlearn things that you've already taught it through human feedback,'' which are not advantageous for many fact-checking applications. \looseness=-1

\subsection{Capability Transparency}\label{sec:transparency}

While participants agreed that GPT-4 outperformed open models, they also said that the capabilities of open models were presented with more \faSearchPlus \textbf{Performance Transparency}. P3 expressed surprise that GPT-4 performed poorly for restructuring their data, a task they thought fell within the model's capabilities, noting that while it looked reasonable ``on the surface\dots when you actually dug into whether it was structured coherently, it wasn't as good.'' Participants also said that \faDownload \ \textbf{Avoiding Deceptive Marketing} motivated use of open models in settings involving user interaction. P17 said the presentation of models like ChatGPT encouraged inappropriate trust by non-experts: ``Yes, there is a popup on ChatGPT, and a line at the ending that tells you the results could be inaccurate, so beware. But how the interface is built, and how it is marketed, how it is presented, and the fact that it answers you in a confident way\dots there is a constant behavioral trick\dots that what you have in front of you, what a machine is telling you, answering to your prompt, is the truth.'' P1 echoed this sentiment, noting that ``There's a problem with people assigning too much credibility to large language models.'' Finally, participants said that providing \faUsers \ \textbf{Tools for Replication} motivated open models. P8 noted that ``we want to give a reader the ability to replicate our research in total\dots every source, everything.'' P9 said that, while all models created problems of explainability, API-gated proprietary models introduced more ``black box problems'' than open models, rendering reproducibility uncertain for end users.

%% file: sections/06_limitations.tex
\begin{figure*}
\small
\centering
\begin{forest}
  for tree={
    align=left,
    font=\sffamily,
    l sep'=8pt,
    s sep = 4pt,
    fork sep'=4pt,
    draw,
  },
  forked edges,
  baseline,
  where level=0{
    tikz={\draw [thick] (.children first) -- (.children last);},
  }{},
  [\textbf{Limitations of Open Models in Fact-Checking}, fill=gray!5
    [\textbf{Performance}, fill=gray!5, align=center
        [\faSearchPlus \ \ Few-Shot Reasoning  \\
        \faSearchPlus \ Multilingual Capabilities \\
        \faDownload \ \ End User Readiness \\
        \faSearchPlus \ Pace of Innovation \\
        \faDatabase \ Average Performance, fill=green!10
    ]    
    ]
    [\textbf{Usability}, fill=gray!5, align=center
        [\faSearchPlus \ \ API Simplicity \\
        \faSearchPlus \ \ Demands of Fine-Tuning, fill=blue!10
        ]
    ]
    [\textbf{Safety}, fill=gray!5, align=center
        [\faDownload \ Handling Charged Input \\
        \faDownload \ Handling User Bias Tests \\
        \faDownload \ Abstaining from False Output \\
        \faDownload \ Ongoing Safety Updates, fill=purple!10
    ]
    ]
    [\textbf{Opportunity Costs}, fill=gray!5, align=center
        [\faUsers \ \ Content Revenue Models \\
        \faUsers \ \ Advertising Models \\
        \faUsers \ \ Brand Name Recognition \\
        \faSearchPlus \ \ Application Ecosystems \\
        \faUsers \ Information Islands, fill=violet!10
    ]
  ]
  ]
\end{forest}

\hfill \break

    \caption{\small A taxonomy of the limitations of open models described by participants as preventing their further adoption in fact-checking.}
    \label{fig:limitations_tree}

\end{figure*}
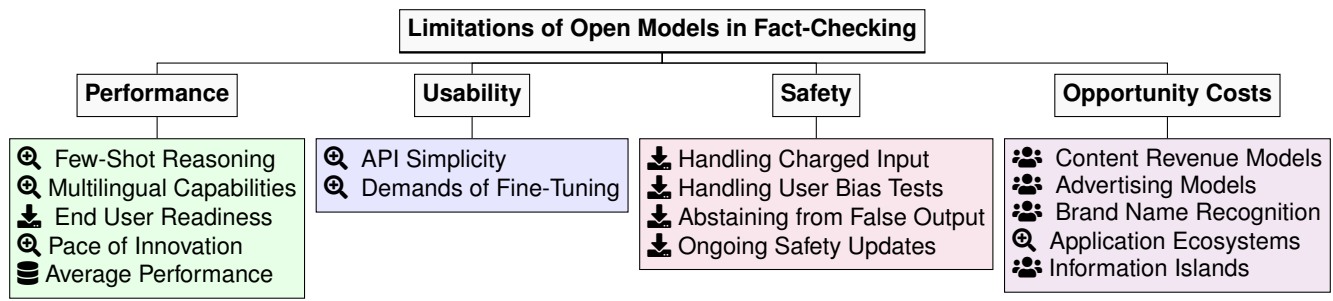

\section{Limitations of Open Models}\label{sec:limitations_open_models}

Despite the many motivations participants gave for preferring open models, most nonetheless used primarily proprietary models, especially OpenAI's GPT models \cite{openaimodels}, citing their Performance, Usability, Safety, and the Opportunity Costs of not participating in proprietary AI ecosystems.

\subsection{Performance}\label{sec:performance}

While fine-tuned open models may achieve the strongest performance on a given task, all participants said that GPT-4 was the best-performing \textit{general-purpose} generative model, motivating its adoption over open models in many settings where general-purpose reasoning is preferable to task specialization. P13 noted that they use GPT-4 over LLaMA models due to \faSearchPlus \ \textbf{Few-Shot Reasoning} disparities, while P9 echoed this in noting that ``if you want to work with a narrative generative AI model, then I think there is, at the moment, no alternative to GPT-4.'' Participants also noted that open models lag OpenAI significantly in \faSearchPlus \ \textbf{Multilingual Performance}. P4, who reported creating primarily non-English models, said that ``we tried the largest [LLaMA models]\dots usually the models tend to perform poorly in languages that are not English. And in the case of OpenAI, it's pretty good.'' P4 also said that performance disparities affected the \faDownload \ \textbf{End User Readiness} of open models: ``when we were developing [our conversational tipline], we tested a bunch of models, especially LLaMA-2, but in terms of performance, it's behind OpenAI significantly\dots I'd love to use open source models.'' P2 said that the \faSearchPlus \ \textbf{Pace of Innovation} in generative AI made it untenable to try to build open models: ``the pace of innovation nowadays is so quick that it's very difficult to keep the pace\dots you don't know which is the new tech that you need to use or which is the model that is going to work\dots even research institutions, they don't know.'' In some cases, participants reported that GPT-3.5-Turbo's inexpensive \faDatabase \ \textbf{Average Performance} was good enough. P1 said that, for retrieving data internally, they needed high recall and not necessarily high accuracy, since human fact-checkers would see the data. They noted ``OpenAI is\dots so easy and cheap. We pay a couple of dollars per month\dots I can use the old (anonymized) servers to run an open language model and see what it can do. But the savings would be minimal, like a couple of dollars, and the results would probably be worse.'' \looseness=-1

\subsection{Usability}\label{sec:usability}

Most participants said the ease of calling proprietary APIs motivated their use of proprietary generative models. P19 said that, even though they prefer to build on open source models, \faSearchPlus \ \textbf{API Simplicity} motivated them to OpenAI: ``OpenAI models are convenient for prototyping because it's just an API call, and you don't have to worry too much about it.'' P2 said, ``what is the good part of OpenAI? Everything with an API is much easier.'' However, P2 also pointed to the possibility that OpenAI's perceived edge in usability is actually due more to its market dominance, noting that ``there are a lot of different frameworks now\dots [such as] LangChain, there are a lot of frameworks that make it easier to work with [open models], and they are the solution.'' Participants noted the \faSearchPlus \ \textbf{Demands of Fine-Tuning} open models drove them to use few-shot proprietary models. Even if task-specific performance exceeding GPT-4 is achievable with fine-tuned open models, data scientists at fact-checking organizations may not have the time  to invest in fine-tuning. P2 said ``the other issue for typical fact-checking organizations, is we are not Microsoft, we are not big technology companies, is the cost of the systems \dots to fine tune your own language model it is too\dots time demanding.'' P4 said it was hard to find time to understand the opportunities of fine-tuning, noting they only used LLaMA ``out of the box, we didn't try to fine tune it. Maybe in the future we will, but\dots resources are limited, so we opted not to explore more.''

\subsection{Safety}\label{sec:safety}

Participants said that proprietary models offer advantages for user-facing applications because of the safety features built into them by larger technology corporations. P13 noted that because user-facing fact-checking technologies typically involved \faDownload \ \textbf{Handling Charged User Input}, reliably adhering to ethical guardrails was essential for maintaining user trust, and open models could not always accomplish this. P4 said that user-facing technologies also had to be prepared for \faDownload \ \textbf{Handling Bias Stress Tests}, noting that users actively attempt to probe their in-beta conversational model (leveraging GPT-4) for political biases: ``one of the most common types of questions that I think people were asking the bot\dots [was] just swapping the name of the [politician] you're asking about\dots and so far we haven't noticed anything that would be concerning for us\dots it's not symmetric, but I think it usually gives you a quite nuanced answer.'' P4 also noted that proprietary models tend to outperform open models in \faDownload \ \textbf{Abstaining from False Output}, noting that users most frequently complained that their RAG-enabled GPT-4 chatbot couldn't answer a question, but that ``in most cases, we didn't have the answer. So the bot behaved as expected as it should, which is to say that it doesn't know the answer to a question instead of trying to come up an invented answer.'' Finally, participants said that proprietary model developers were better positioned to perform \faDownload \ \textbf{Ongoing Safety Updates}. P21 said that ``fact check organizations are underequipped to really test everything and keep up with everything on an ongoing basis''; accomplishing this with open models, they said, would require ``revenue streams\dots that's still trying to be figured out.'' \looseness=-1

\subsection{Opportunity Costs}\label{sec:opportunity}

Participants worried that using open models would entail foregone opportunities for integration into an emerging information ecosystem, potentially lessening the relevance of their content and precluding them from taking advantage of new streams of revenue. P9 said that \faUsers \ \textbf{Content Revenue Models} would ``totally define how we will work with Gen AI models, and if we will work in close partnerships with these companies, or if we will work with open source models, or if they will be state funded projects. All the journalistic companies say we need it to save democracy, and we push for open source models; or we say, no, we're fine, we're getting millions and millions from Google, and OpenAI, and Amazon, and so on. So yeah, we're sort of at a crossroads.'' P21 noted that, in addition to direct compensation for content, proprietary models might be preferred due to \faUsers \ \textbf{Advertising Models}, noting that ``the value exchange that we're familiar with today with Google search is you structure your metadata the right way, and you'll show up in Search, and you'll get clicked on, and you'll generate ad revenue. And perhaps Google's even the one who buys those ads, and the revenue comes from them anyway. I think that [Google is] more ready as an organization to think about paying publishers. With ChatGPT that's not really established yet.'' P21 also said that they hoped to leverage the \faUsers \ \textbf{Brand Name Recognition} of emerging AI ecosystems to increase their reach, noting ``ultimately ChatGPT is the biggest brand name in chatbots, and they are also already integrated and backing Bing\dots to reach all of the users who are using chatbots, it's not realistic to think that we'll make the biggest splash just by having our own private code base and onsite chat experience. I do think that we have to be able to play into the bigger arena.'' P10 further discussed the possibility of integrating their internal models into broader \faSearchPlus \ \textbf{Application Ecosystems}, noting that ``we use the Google workspace workflow, so it really helps to incorporate the Bard features.'' Finally, P21 raised broader concerns about the information ecosystem, describing the possibility of \faUsers \ \textbf{Information Islands} if internet users privately interact with open chat models: ``if people choose to search or engage with a chat bot that\dots isn't trained on your content, we stand to have these kind of information islands\dots someone who searches on Google today or Bing today, they're going to get [our content] as results that were fully enabled to be crawled by search engines. But as more and more chat bots emerge\dots that is kind of a threat, I think, to information integrity overall.''

%% file: sections/07_discussion.tex
\begin{table*}[]
    \centering
\fontsize{7.5}{8}\selectfont
    \begin{tabular}{|p{2.8cm}|p{6.5cm}|p{7cm}|}
    \toprule
    \multicolumn{3}{|c|}{Research Directions for Use of \textbf{Open Models} in Fact Checking} \\
    \hline
     Concern & Research Question & Research Directions \\
     \midrule
      \textbf{Performance} of Open Models & How can open models offer competitive performance to proprietary models while maintaining an approachable conversational interface? & Developing scale-efficient open models; Developing more suitable evaluation suites for fact-checking; Developing evaluations specifically for open models. \\
     \hline
      \textbf{Usability} of Open Models & What kinds of open application interfaces can help fact-checkers feel as comfortable with open models as proprietary models for inference and fine-tuning? & Creating open source or public APIs of comparable simplicity to OpenAI; Decreasing the time and expertise demands of fine-tuning.  \\
      \hline
      \textbf{Safety} of Open Models & How can open models achieve the actual and perceived safety of proprietary models without incurring significant time and cost burdens? & Community standards for lightweight red-teaming of open models; Technologies to monitor the fairness of model responses in user-facing conversational tiplines. \\
      \hline
      \textbf{Opportunity Costs} of Open Models & How can open model ecosystems facilitate reliable revenue streams for fact-checking organizations similar to those in proprietary model ecosystems? & Fostering community and collaboration between fact-checking organizations and AI developers; Developing a revenue model for open model ecosystems. \\
      \hline
    \multicolumn{3}{|c|}{Research Directions for Use of \textbf{Proprietary Models} in Fact Checking} \\
    \hline
    Lack of \textbf{Autonomy} & How can proprietary model developers assure clients of access to models integrated in data pipelines?  & Developing approaches to allowing selective access to deprecated models. \\
    \hline
    Lack of \textbf{Data Privacy and Ownership} & How can proprietary model developers guarantee clients that their data won't be used inappropriately? How can they guarantee clients that using their products will not put them in legal jeopardy? & Developing models of compensation for publishers; Supporting legal standards for client data privacy; Clearly communicating when data will be retained or used outside of initial context. \\
    \hline
    Lack of \textbf{Application Specificity} & How can developers afford clients more control over tone in user-facing applications and more contextually appropriate means of processing toxic content? & Developing approaches for personalized models that meet fact-checking use cases without compromising model safety. \\
    \hline
    Lack of \textbf{Transparency} & How can developers of API-gated proprietary models provide transparency about their capabilities and afford fact-checking organizations the ability to explain their use to audiences? & Communicating capabilities of domain-specific GPTs through systematic quantitative and qualitative evaluation; Providing more explainable outputs such as token-level probabilities. \\
    \hline
    \end{tabular}
    \caption{\small We provide research directions for addressing limitations of both open and proprietary models proprietary models in fact-checking.}
    \label{tab:research_directions}
\end{table*}

\section{Discussion}

Our findings add new perspective to the evolving conversation about the use of open or proprietary generative AI, surfacing the tradeoffs that face \textit{organizations} as they consider whether and where to adopt open models. Some tradeoffs would seem contradictory without being contextualized within the fact-checking data science pipeline. For example, participants chafed at limitations imposed by toxicity guardrails when using OpenAI models for internal analysis tasks, yet reported \textit{relying} on those guardrails for user-facing applications. Moreover, while participants preferred fine-tuned open models for strong performance on internal tasks, they also accepted ``good enough'' performance via the cheap, easy-to-use GPT-3.5-Turbo API for high-recall data retrieval tasks, over the time cost incurred for an open model.

Table \ref{tab:research_directions} outlines research questions and directions arising from this work, highlighting the need to address issues with both open and proprietary models. Fact-checking organizations appear at first to face diverging futures: one in which they adopt open models and control their AI infrastructure; and another operating profitably in the emerging ecosystems of chat-driven interfaces now both collecting and delivering information in interactions with end users. However, our findings surface that the future may resemble the present, with organizations employing specialized open models for mission-critical internal tasks, proprietary generalist models to handle interactions with end users, and the least expensive option when performance just needs to be good enough. Yet even in this blended future, proprietary models face challenges around data ownership. Participants in our study consistently voiced unease with their work and that of their colleagues being used without consent to create a lucrative competing product. Some participants also avoided generative AI for fear of inadvertently committing plagiarism. These concerns echo those of many journalists, as well as domains like visual art, wherein artists find their work devalued by a tool trained on their own content \cite{jiang2023ai}. Addressing issues of data privacy, ownership, and compensation may prove paramount for proprietary model providers to establish relationships with fact-checking organizations and other producers of factual and creative content.

\subsection{Limitations and Future Work}

While we sought to capture perspectives of fact-checking organizations globally, our participants all spoke English, and most used models in English. Aside from noting that OpenAI's models outperform open models in multilingual settings for generalist conversational tasks, our work does not speak to the additional intricacies of multilingual NLP, including NLP in low-resource languages. Additionally, our work is concerned with open generative models, rather than with all open models (such as autoencoder transformers models). Future work might study perspectives on open models and proprietary models of any architecture.

%% file: sections/08_conclusion.tex
\section{Conclusion}

We contribute an organizational perspective on open and proprietary models, finding that while fact-checking organizations prefer open models for Organizational Autonomy, Data Privacy and Ownership, Application Specificity, and Capability Transparency, they nonetheless use proprietary models for their Performance, Usability, and Safety. We suggest open models could benefit from increased usability, while proprietary model providers must address concerns over data ownership and compensation. As generative AI ecosystems become more mature, financial models may determine whether fact-checking organizations disseminate information in proprietary ecosystems, or form independent alternatives.

\section{Researcher Positionality}

The authors are researchers at a university in the United States. The second author has long-standing relationships with fact-checking organizations around the world, which they have fostered throughout their career. The first author has had informal experience working with fact-checking organizations in research contexts, and has technical expertise with generative AI and related NLP and computer vision technologies. We sought to include a diversity of perspectives in this research by interviewing individuals working at fact-checking organizations around the world; nonetheless, we note that our positionality as researchers located in the U.S. prevents us from comprehensively describing the contexts in which our diverse pool of participants operate.

\section{Adverse Impacts}

We sought in this research to provide a balanced perspective representative of the views of our participants. However, in suggesting a research agenda that accounts for both open and proprietary models, our study may seem to place these technologies on equal footing. The harms arising from proprietary models are likely to be very different than those arising from open models, and potentially greater in scale due to their much large end user base at the present time. Thus, this research should not be read as concluding that open and proprietary models have \textit{similar} societal implications, but rather that challenges faced by these models are both sufficiently important as to warrant concurrent research agendas. Moreover, we have focused not on the question of \textit{whether} use of generative AI is desirable in fact-checking organizations, and what the drawbacks of that might be, but on the tradeoffs of open and proprietary models. We note that there may be notable harms in use of any form of generative AI in fact-checking, particularly when the technology is positioned as a replacement for human expertise \cite{juneja2022human}.

\section{Ethical Considerations}

We centered the role played by experts at fact-checking organizations, as well as the data pipelines employed by these organizations. However, we note that there are many human roles that we have not managed to surface in this work, including roles held by individuals who participate in data pipelines in ways that would not traditionally be recognized as data science \cite{muller2019human}. This is a consequence of our research questions, which focus on uses of open and proprietary language models, rather than on the less recognized data labor performed by individuals and organizations that seek to counter misinformation. Future work might more directly address these less obvious \textit{human} contributions to the data pipelines surfaced in the present work